# Performance and Detection of *M*-ary Frequency Shift Keying in Triple Layer Wireless Sensor Network

Mohammad Waris Abdullah[1], Nazar Waheed[2]

Dept of Communications & Network Engineering
King Khalid University, Abha, Saudi Arabia
`abdullah@kku.edu.sa`[1], `nazar.waheed@gmail.com`[2]

*ABSTRACT*

*This paper proposes an innovative triple layer Wireless Sensor Network (WSN) system, which monitors M-ary events like temperature, pressure, humidity, etc. with the help of geographically distributed sensors. The sensors convey signals to the fusion centre using M-ary Frequency Shift Keying (MFSK) modulation scheme over independent Rayleigh fading channels. At the fusion centre, detection takes place with the help of Selection Combining (SC) diversity scheme, which assures a simple and economical receiver circuitry. With the aid of various simulations, the performance and efficacy of the system has been analyzed by varying modulation levels, number of local sensors and probability of correct detection by the sensors. The study endeavors to prove that triple layer WSN system is an economical and dependable system capable of correct detection of M-ary events by integrating frequency diversity together with antenna diversity.*

## 1. INTRODUCTION

The development of Wireless Sensor Networks (WSNs) marks a significant departure from the conventional method of decentralized detection by wired sensors. The wireless sensors are capable of transmitting information over a common wireless spectrum. Moreover, recent advances in wireless communications and electronics coupled with an urgent need for the development of low cost, low power and multifunctional sensors have motivated a number of researches worldwide.

Wireless Sensor Networks (WSNs) come with an inherent advantage of easy deployment, which is a very appealing attribute for battlefield surveillance, environmental monitoring, etc.[1] However, the presence of highly dynamic RF channels demands for the design of high efficiency detection algorithms. These sub-optimum and optimum detection algorithms include Neyman- Pearson detection [2,9], Bayes Detection [3,10], Maximum likelihood detection [2-7], Maximal Ratio Combining and Equal Gain Combining [3-6, 9], Chair- Varshney Fusion detection [3-6, 11], etc. In [9], the use of Maximal Ratio Combining (MRC), Equal Gain combining (EGC) and Selection Combining (SC), has been detailed with their application pertinent to diversity combining schemes for *M*-ary symbols.

Hahn [13] and Lindsey [14] were the first to consider the use of Equal Gain combining for MFSK detection using square law combining over Rayleigh and Rician Fading channels. They assumed that MFSK signals are disturbed independently by Rayleigh fading and Additive White Gaussian Noise. The method used for non-coherent detection in this research is quite similar to the approach of Hahn. The difference is related to the approach of detection and the extent of diversity being employed by this research. Unlike Hahn, in this research, all the sensors make independent detection and transmission and the detection technique is sub-optimal. A more general result for Nakagami-m channel was obtained by Crepeau [15] for the case of no diversity and by Weng and Leung [16] for the case of square- law diversity combining. The average Bit Error Rate (BER) performance of non-coherent orthogonal MFSK modulation scheme in Nakagami channels was also analyzed by Yan Xin et al [17]. It was observed that sub-optimum receiver suffers from non-coherent combining loss at sufficiently low Signal to Noise Ratios (SNRs). Various researches have been conducted to determine '*closed-form mathematical expressions*' for error performance of MFSK signals with diversity reception. Proakis [12] developed a general mathematical formula for evaluating the error rate for multi-channel non-coherent and differentially coherent reception of binary signals over $L$ independent Additive White Gaussian Noise (AWGN) channels. Similar analysis was carried out by Marvin and Alouini [9] where they used EGC combining scheme to determine the BER performance of non-coherent M-ary FSK over AWGN and Rayleigh Fading channels. The error performance analysis of coherent MFSK using MRC diversity scheme was probably first attempted by Al- Hussaini et al [18] and later by Aalo [19]. Their work was mainly confined to detection of binary alphabets in Nakagami-m fading channels. Dong [20], Beaulieu [21] gave closed form mathematical expression for Symbol Error Probability (SEP) for coherent MFSK in Rayleigh fading channels with $M \leq 4$. Later Xiao and Dong did the same analysis for a more general result in Nakagami-m fading channels [22] and Rician channels [23]. These expression involved infinite integral limits and infinite series, making them very difficult to solve. Marvin and Alouni [9] tried to simplify these expressions by using exponential type integrals with finite limits. Later, Paris et al [24] evaluated the Bit Error Probability (BEP) of coherent MFSK in Rician channels with MRC.

This research endeavors to carry forward the work done in the performance analysis of the detection of MFSK and extend it to the domain of Wireless Sensor Networks (WSNs). The proposed system is a parallel triple layer network model [11], which performs non-coherent detection of $M$-ary states of any single event. The source event has $M$ possible states, which are detected by the local nodes in the form of $M$-ary amplitude shift keying (MASK) signals. The information from the sensor is communicated to the fusion centre using $M$-ary Frequency Shift Keying (MFSK) [9] modulated symbols over Rayleigh Fading Channels in separate time slots. The choice of Rayleigh fading channel can be attributed to the fact there under many scenarios there might not be a line of sight connection between the sensors and the fusion layer.

The remaining paper is organized as follows. Section II is devoted to system description. In this section, the source event, local sensor processing and the functioning of the fusion centre is discussed in greater detail. Further, mathematical relationships together with the proposed receiver structure are also explained. Section III provides the simulation results and corresponding discussions. The simulations are carried out with the help of MATLAB. Finally, section IV provides the derived conclusion of the research.

## 2. System Description

In this section, all the three stages of the system as shown in figure 1 are described extensively. The MFSK WSN system under investigation is a triple layer wireless sensor network system as illustrated in figure 1. The first layer being of local sensors distributed over a geographical area, the second layer is the channel and the third layer consists of the fusion centre. The first layer consists of $L$ sensors and all of them concurrently observe an event. The main function of the first stage of the system is to perform quantization of any continuous event like temperature, pressure, humidity, etc. into $M$ states of $M$-ary Amplitude Shift Keying (MASK). These quantized states are then directly mapped to the MFSK symbols. All this processing takes place at the first level of the system, i.e. at the sensors. The sensors then send the observed values to the fusion centre over Rayleigh Fading channels in separate time slots. The choice for Rayleigh fading channel indicates that there might not be a line of sight path connection between the sensors and the fusion layer. This channel constitutes the second layer of the system. The final layer is the fusion centre. The fusion centre determines the state of the source event with the help of a non-coherent receiver. The detection at the receiver is based on Selection Combining (MRC), which unlike Equal Gain Combining (EGC) and Maximal Ratio Combining (MRC) does not require any channel information, like signal phase, channel amplitude, etc. The receiver circuitry is therefore simple as it just processes one of the channels. The receiver at the fusion centre chooses the branch with the highest Signal to Noise Ratio (SNR) [10].

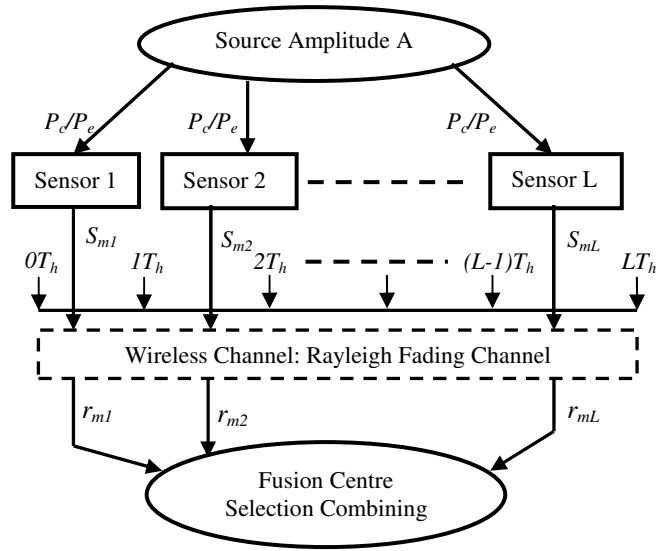

**Fig. 1** MFSK Triple Layer Wireless Sensor Network with $L$ nodes for detecting $M$-ary events. The sensor detects an event with probability of correct detection ($P_c$) and transmits the detected event in a particular time slot ($T_h$) [8].

### 2.1 System Event

The system is designed for detection of single events like temperature, pressure, humidity, etc. The detection of an event takes place in the presence of Additive White Gaussian Noise (AWGN). After detection, the next stage is quantization where an observed source event A is quantized to any one of the $M$ modulation levels of $M$-ary ASK. These quantized values are then directly mapped to one of the $M$ modulation levels of MFSK. Since, the detection at the fusion centre is non-coherent in nature, therefore for a given MFSK symbol belonging to the set, [*1, 2, 3…..M*], the minimum frequency separation between the modulated symbols is kept at $1/T_s$, where $T_s$ is the symbol duration. Consequently, the corresponding frequency set becomes [*1/ $T_s$, 2/$T_s$, 3/ $T_s$ …..M/ $T_s$*] [9].

## 2.2 Sensor Processing

In one symbol duration, the $l^{th}$ sensor is detecting one event and therefore the transmitted symbol is represented as [9]

$$S_{ml} = \sqrt{P}\, exp\, [j\{2\pi(f_c + f_{ml})t + \emptyset_l\}] \quad (1)$$

The source state $S_{ml}$ (MFSK symbol) is determined by two factors *m* and *l*. The factor *m* corresponds to one of the *M* states and *l* corresponds to one of the sensors, ranging from [*1, 2, 3……L*]. The transmission power per symbol, which is denoted as *P* is same for all the sensors. The carrier frequency is $f_c$. The frequency for each symbol *m* transmitted by the $l^{th}$ sensor is represented as $f_{ml}$. The frequency $f_{ml}$ belongs to the set given by [*1/Ts, 2/Ts, 3/Ts …..M/Ts*]. The variable $\Phi_l$ is the initial phase introduced due to carrier modulation, with respect to each $l^{th}$ sensor.

The system uses a scheme where each sensor transmits one symbol in a given time slot, $T_h$, which is equal to $T_s/L$. This avoids inter-symbol interference. Therefore, the signal given by (1) transmitted by $l^{th}$ sensor during the time slot given by $iT_s < t \leq (i+1)T_s$ is expressed as

$$S_{ml} = \sqrt{P}\, \rho_{th}(t - iT_s - [l-1]T_h) \times exp\, [j\{2\pi(f_c + f_{ml})t + \Phi_l\}] \quad (2)$$

The factor $\rho_{th}$ denotes the Gaussian pulse shaped signaling waveform, which is defined over the time slot [*0, $T_h$*]

## 2.3 Fusion Centre Processing

Each sensor would transmit the symbol to the fusion centre over a Rayleigh fading channel and therefore the received symbol $r_{ml}(t)$ at the fusion centre is expressed as follows

$$r_{ml}(t) = h_l S_{ml}(t) + n(t) \quad (3)$$

Where $h_l = \alpha_l\, exp(j\theta_l)$ is the channel gain with respect to the $m^{th}$ symbol and the $l^{th}$ sensor, which is assumed to be constant over one symbol duration, $T_s$. The transmitted symbols $r_{ml}$ received at the fusion centre is detected with the help of a receiver based on selection combining diversity scheme. The system model of detection scheme uses a comparator that determines the Signal to Noise Ratio (SNR) of each branch and chooses that branch, which has the maximum $r^2_{ml}+N_l$. The noise power $N_l = N$ is same for all branches [10]. Further, it is assumed that there is sufficient antenna spacing and therefore the branches are independent of each other. The decision variables are determined as

$$Z_M = max\, [r^2_{ml} + N], where\, m \in [1,2,3,…..M]\, and\, l \in [1,2,3….L] \quad (4)$$

Now, the largest value from the set of [$Z_1, Z_2.....Z_M$] is selected and mapped to an integer in the range of [*1,2,3,....M*] to determine the event observed by the *L* sensors [9].

## 3. Simulation Results

In this section, the error performance of MFSK WSN is simulated and analyzed for single event detection and the sensors are only affected by Additive White Gaussian Noise (AWGN). The channels from the local sensors to the fusion centre are independent Rayleigh Fading channels. The Bit Error Rate Performance (BER) is represented and investigated to understand the performance of the proposed system for a convenient comparison with other conventional systems. The proposed system is made to operate over Rayleigh Fading channels under different conditions. These conditions include, increasing number of modulation levels, increasing number of local sensors and varying probability of correction detection of an event by the sensor.

### 3.1 Increasing Number of Sensors

In figure 2, it is observed that as the number of local sensors increases from 2 to 30, the error performance degrades and the degradation is more prominent at lower SNR values. This feature can be attributed to the fact that as the number of sensors increases, the transmitted power per sensor decreases. Further, the diversity gain achieved through geographically distributed sensors is also lost because the separation between the nodes decreases [9]. At low SNR values up to 10 dB, error floors are present for all the systems and any increase in SNR results in crossover of error curves.

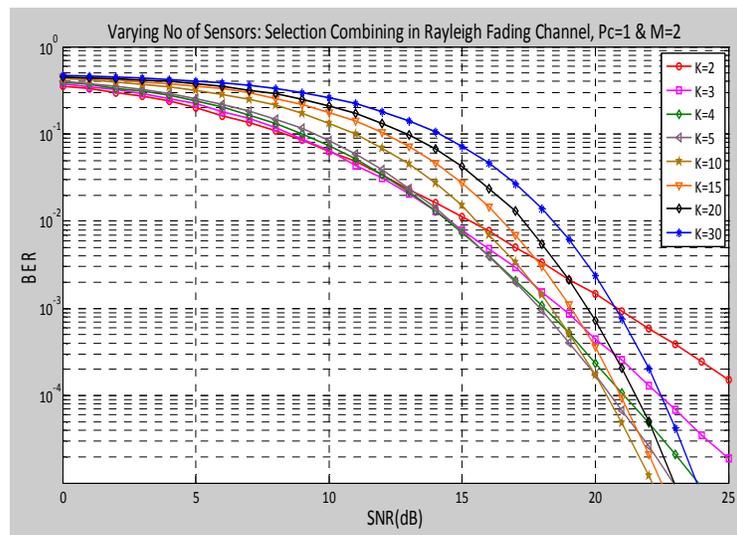

**Fig 2:** Bit Error Rate (BER) vs Signal to Noise Ratio (SNR) performance of 2-FSK WSN for different number of sensors over Rayleigh Fading Channels with probability of correct detection equal to unity.

However, as the SNR values increases above 20 dB, it is observed that systems with more number of sensors start giving better performance. *However, a limiting value is again encountered!* As the number of sensors increases beyond 10, the BER performance again starts degrading. A system with 10 sensors is able to achieve an error rate of $10^{-5}$ at 22.5 dB while another system with 15 sensors needs an SNR of 23 dB for the same performance. It can be therefore safely assumed that if the detection scheme is Selection Combining (SC), then increasing the number of sensors beyond a certain number would be futile. This is because selection combining is not at an optimal technique. For a better understanding of this feature, the effect of increasing the number of sensors at a higher modulation levels is further analyzed in figures 3-4.

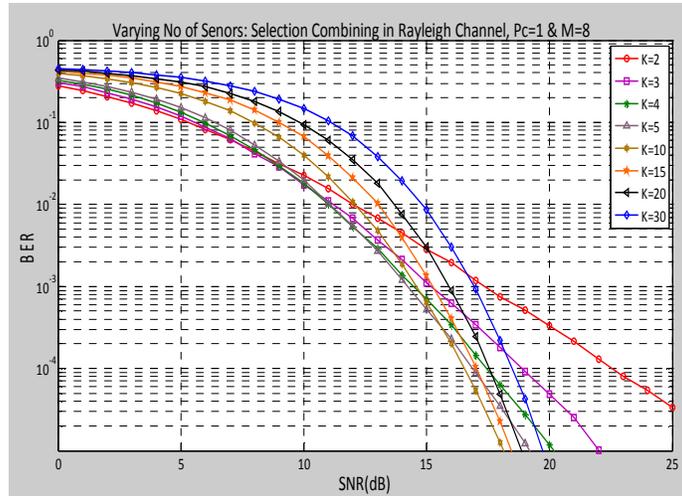

**Fig 3:** Bit Error Rate (BER) vs Signal to Noise Ratio (SNR) performance of 8-FSK WSN for different number of sensors over Rayleigh Fading Channels with probability of correct detection equal to unity.

The 8-FSK system shows better error performance in comparison to a 2-FSK based WSN system. As evident from Figure 3, for the same BER of $10^{-5}$, 8-FSK WSN gives an improvement of almost 5dB in comparison to a 2-FSK for a 10 sensors WSN system. Although, the performance improves for 8-FSK, but still the system with higher number of sensors have more error rate. The improvement in error performance for 8-FSK modulation scheme motivated the study to do the same analysis for still higher modulation levels.

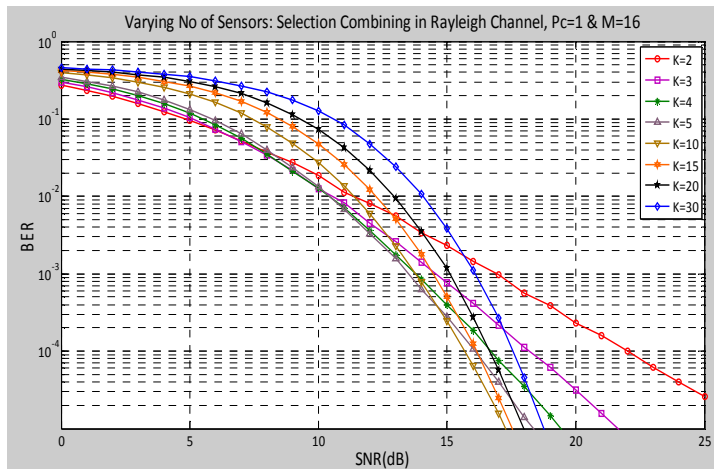

**Fig 4:** Bit Error Rate (BER) vs Signal to Noise Ratio (SNR) performance of 16-FSK WSN for different number of sensors over Rayleigh Fading Channels with probability of correct detection equal to unity.

It is observed from Figure 4 that for 16-FSK, the error performance shows further improvement in comparison to 8-FSK. It should be noted that in the above simulation results (Figure 2-4), it is assumed that all the sensors detect an event with probability of correct detection equal to unity. However, this is not always possible in a dynamic environment. The situation can be further aggravated by the presence of faulty sensors. The research therefore makes an attempt to analyze the performance of MFSK WSN in a scenario where the probability of correct detection by the local sensor nodes is less than unity.

### 3.2 Varying Probability of Correct Detection ($P_c$)

The study simulates the scenario where the sensors are not able to detect an event correctly.

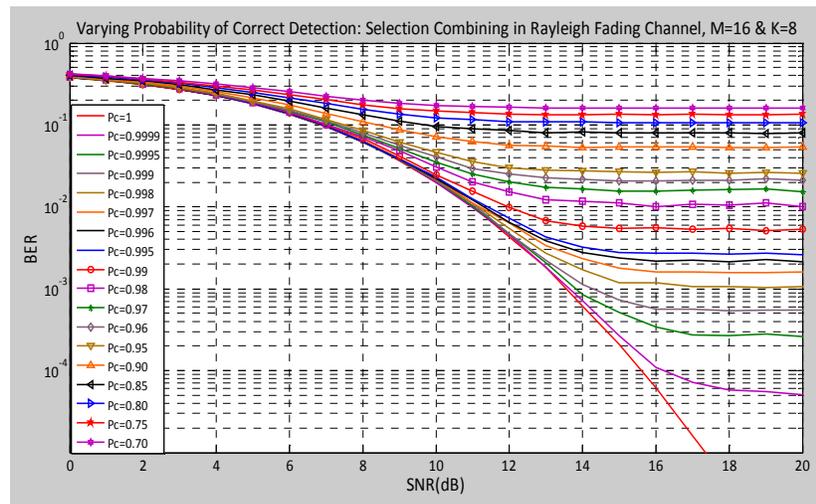

**Figure 5:** Bit Error Rate (BER) vs Signal to Noise Ratio (SNR) performance of 16-FSK 8 node WSN system over Rayleigh Fading Channels for probabilities of correct detection less than unity.

It is observed in Figure 5 that as the probability of correct detection decreases from 1, the performance of the WSN system starts degrading. The poor is the value of probability of correction detection, the greater is the variation from the ideal performance. Even if the sensor's probability of correct detection falls to $P_c=0.9995$, the error rate increases from 6.197e-05 to 3.407e-04. It can be concluded that the performance of SC based receiver at $P_c< 1$ is worst than Maximal Ratio Combining (MRC) and Equal Gain Combining (EGC) based detection systems. [25-26]

The BER performance improves as the SNR increases. This is because at higher SNR, the sensors are able to transmit at an increased power, as a result the probability of a signal to remain in deep fade decreases. For e.g. at $P_c=0.9995$, the BER is 8.55 e-04 at SNR=14 dB, which improves to 3.407 e-04 at SNR=16 dB and further improves to 2.656 e-04 at SNR=18 dB. However, for lower values of $P_c$, error floors are observed, which stay unaffected by any increase in transmission power.

There are now two options available to improve the performance of the system in an environment where the sensors are faulty or are not able to detect an event correctly. The first approach is to increase the number of sensors and the second approach is to increase the number of modulation levels. The research has simulated and analyzed both the options.

### 3.3 Increasing Number of Sensor when $P_c <1$

In the preceding section, it is observed that as the value of $P_c$ decreases from 1, the error performance starts degrading. Therefore, it was envisaged that the performance of the system can be improved if the number of sensors is increased. This section attempts to simulate and analyze the error performance of 16-FSK WSN system by increasing the number of sensors. Here, each sensor is detecting an event with a probability of correct detection equal to 0.999.

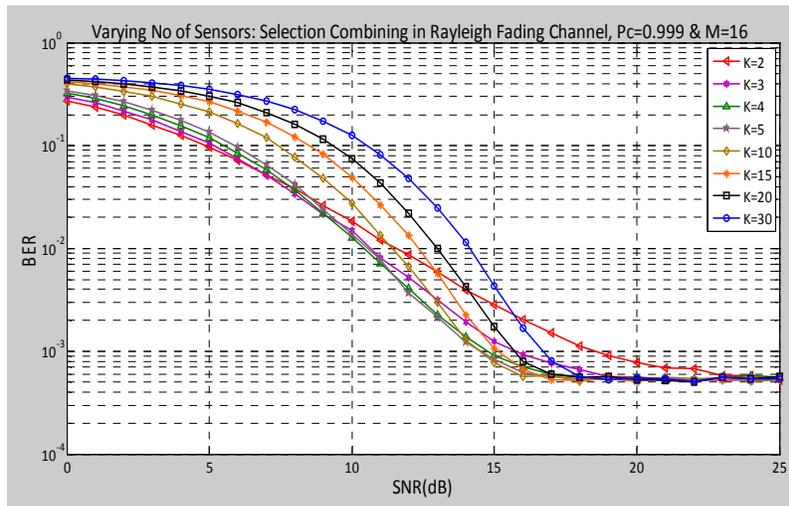

**Fig 6:** Bit Error Rate (BER) vs Signal to Noise Ratio (SNR) performance of 16-FSK WSN system over Rayleigh Fading Channels for probabilities of correct detection equal to 0.999

It is observed that as the number of sensor nodes is increased, the error performance of the WSN degrades, for the reasons explained before. For low values of SNR, the error rate keeps on degrading as the number of sensors is increased. However, for higher SNR values, the error rate does not show much variation (stays at approximately 5e-04), irrelevant of the number of sensors used in the system. The next available option to improve performance is to increase the number of modulation levels. This feature is simulated and analyzed in figure 10.

## 3.4 Increasing Number of Modulation Levels

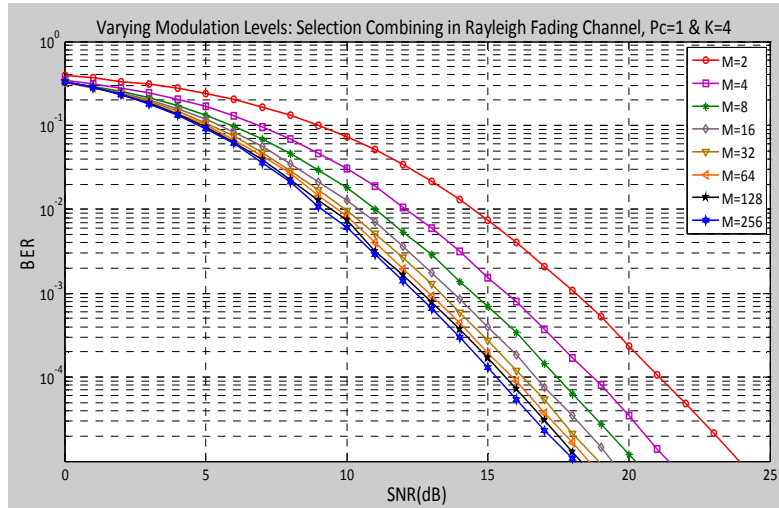

**Fig 7:** Bit Error Rate (BER) vs Signal to Noise Ratio (SNR) performance of MFSK for 4 nodes WSN system over Rayleigh Fading Channels for different modulation levels with probability of correct detection equal to unity.

It is illustrated in figures 2-6 that systems using more number of sensors at low SNRs are giving poor performance. Therefore, it becomes very obvious to study the error performance of the WSN system using higher modulation levels. In figure 7, it is observed that as the modulation level increases from 2 to 256, the performance of the system improves. At a lower SNR value for e.g. 5 dB, a major improvement in BER is observed as we progress from 2FSK to 256FSK. The margin of error rate improvement in this case is from 0.2414 to 0.09148 (almost 62.5%), which keeps on improving as the transmitted power of the sensors is further increased. The above analysis is carried out for 4 nodes WSN system.

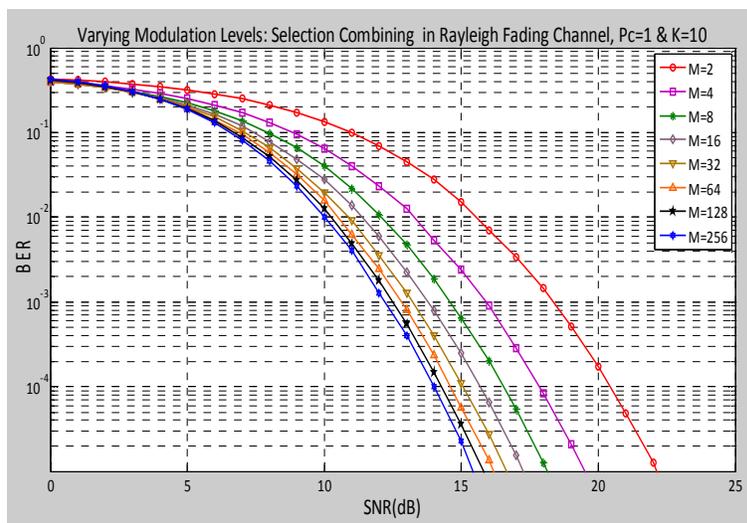

**Fig 8:** Bit Error Rate (BER) vs Signal to Noise Ratio (SNR) performance of MFSK for 10 nodes WSN system over Rayleigh Fading Channels for different modulation levels with probability of correct detection equal to unity.

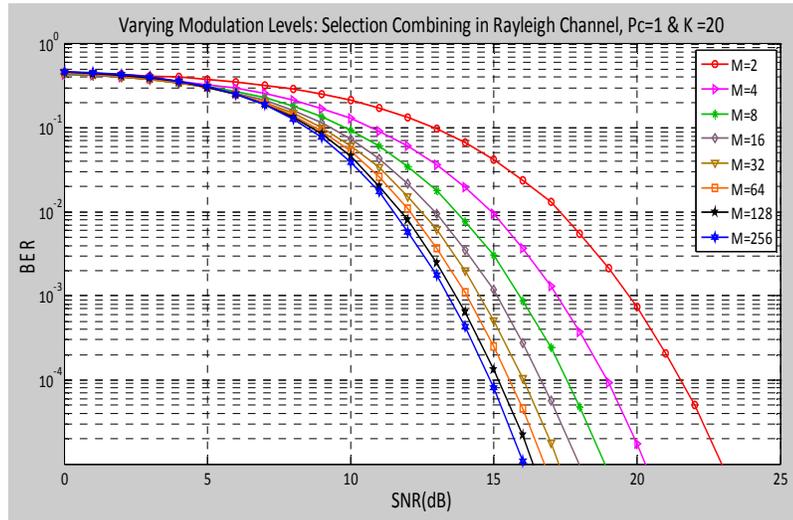

**Fig 9:** Bit Error Rate (BER) vs Signal to Noise Ratio (SNR) performance of MFSK for 20 nodes WSN system over Rayleigh Fading Channels for different modulation levels with probability of correct detection equal to unity.

Similar analysis and simulations are carried for a network with 10 and 20 sensors each, as in Figure 8-9. However, it should be pointed out that the performance is still poor in comparison to MRC and EGC based detection schemes [25-26]. This can be attributed to the fact that 'Selection Combining' is not using all the information available from other sensors. The improvement in error rate is because the information carried per channel increases when higher modulation levels are used. This is indeed an interesting feature of '*a not so very bandwidth efficient scheme*', which is in contrast to other modulation schemes such as MQAM or MPSK. In figures 7-9, it is depicted that as the modulation levels increases from M=2 to 256, the error performance improves. The number of sensors used for each analysis is 4, 10 and 20 respectively. However, as the number of sensors is increased from 10 to 20, the error performance again degrades. This can be attributed to the fact that with an increase in the number of sensors, the transmission power available per sensor decreases. Moreover, the channels from the sensor to the fusion centre are no more uncorrelated or independent.

## 3.5 Increasing Number of Modulation Levels when $P_c < 1$

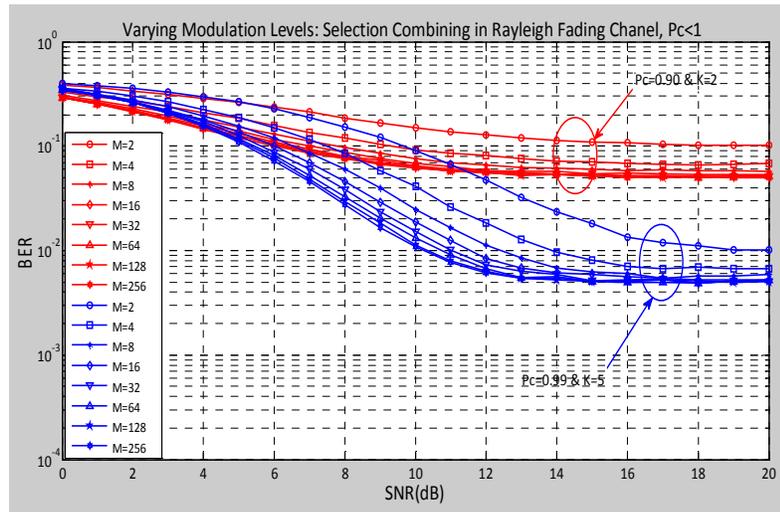

**Fig 10:** Bit Error Rate (BER) vs Signal to Noise Ratio (SNR) performance of MFSK WSN system for different modulation levels for probabilities of correct detection equal to 0.9 and 0.99.

Figure 10, shows that the performance of a WSN system depends upon three major parameters, number of sensors in the system, modulation level and probability of correct detection of an event by the sensors. As shown in the figure above, the BER performance improves as we move from lower value of $P_c=0.9$ to a higher value of $P_c= 0.99$, together with an increase in the number of sensors from 2 to 5. For $P_c= 0.99$ and SNR= 10dB, the BER for 2-FSK is 0.09034, which improves to 0.04123 for 4-FSK. The BER keeps on improving and becomes 0.02478 for 8-FSK, 0.01878 for 16-FSK, 0.01523 for 32 FSK. However for higher modulation schemes, the error rate remains mostly in the range of 0.005 (approximately) i.e. not much improvement is observed even when the SNR value is increased above 12 dB.

## 4. Conclusion

The study proposes an innovative triple layer WSN system, which allows the local sensor nodes to communicate with the fusion centre using MFSK modulation scheme over independent uncorrelated Rayleigh fading channels. As a result, the system enjoys both spatial as well as frequency diversity. The MFSK modulated symbols are detected at the fusion centre with the help of Selection Combining based non coherent detection. The detection scheme is sub-optimal but it provides the benefits of a simple and economical receiver circuitry. The study has concluded that increasing the number of sensors to improve error performance beyond a certain number is a futile option. Further, at lower SNR, the system with lesser number of nodes provides better results than those with a higher number. It is only at an increased transmission power, that the system with more number of sensors starts giving better performance. This phenomenon is attributed to the fact that as the number of sensor nodes increases the separation between them decreases and as a result the fading channels are no more independent. The study also shows that as we move towards higher modulation levels, the error performance of the system improves. However, this improvement is limited to a certain number of modulation levels only. At lower SNR values, the error performance doesn't show much improvement even

International Journal of Computer Networks & Communications (IJCNC) Vol.4, No.4, July 2012if the system uses larger number of sensors. This can be attributed to the fact that SC is not using all the information available from other sensors. The study demonstrates that the overall performance of the system is extremely dependent on the capability of the sensors to correctly detect an event. Even a small deviation harms the overall efficiency of the system. The study concludes that in an environment where the sensors are not able to detect correctly, increasing the number of modulation levels together with increasing the number of sensors up to a certain level would improve the performance of the system.

## 5. References

[1]     A. Swami, Q. Zhao, Y.-W. Hong, & L. Tong. (2007). *Wireless Sensor Networks: Signal Processing and Communications Perspectives*. Chichester, England: John Wiley & Sons.

[2]     R. Viswanathan & P. Varshney. (1997).  Distributed detection with multiple sensors: Part i- Fundamentals,  *Proceedings of the IEEE*, vol. 85, no. 1, pp. 54 – 63.

[3]     B. Chen, R. Jiang, T. Kasetkasem, & P. Varshney. (2004).  Channel aware decision in wireless sensor networks,  *IEEE Transactions on Signal Processing*, vol. 52, no. 12, pp. 3454 – 3458.

[4]     R. Niu, B. Chen, & P. Varshney. (2006).  Fusion of decisions transmitted over Rayleigh fading channels in wireless sensor networks,  *IEEE Transactions on Signal Processing*, vol. 54, no. 3, pp. 1018 – 1027.

[5]     B. Chen, L. Tong, & P. Varshney. (2006).  Channel-aware distributed detection in wireless sensor networks,  *IEEE Signal Processing Magazine*, no. 3, pp. 16 – 26.

[6]     Y. Lin, B. Chen, & P. Varshney. (2005).  Decision fusion rules in multihop wireless sensor networks,  *IEEE Transactions on Aerospace and Electronic Systems*, vol. 41, no. 2, pp. 475 – 487.

[7]     R. Jiang & B. Chen. (2005).  Fusion of censored decisions in wireless sensor networks,  *IEEE Transactions on Wireless Communications*, vol. 4, no. 6, pp. 2668 – 2673.

[8]     Fucheng Yang, Lie-Liang Yang, Huangfu Wei & Limin Sun. (2010).  Frequency-hopping/M-ary frequency-shift keying for wireless sensor networks: Noncoherent detection and performance,  *Wireless Communication Systems (ISWCS). 2010 7th International Symposium on* , pp.135-139.

[9]     M. K. Simon & M. S. Alouini. (2000). *Digital Communications Over Fading Channels*. New York: Wiley.

[10]   A. Goldsmith. (2005). *Wireless Communications*. Cambridge, England: Cambridge University Press.

[11]   R.Niu, B. Chen & P.K. Varshney. (2006).  Fusion of Decision Transmitted Over Rayleigh Fading Channels in Wireless Sensor Networks,  *IEEE Transactions on Signal Processing,* vol. 54, pp.1018-1027.

[12]   J.G. Proakis. (1995). *Digital Communications,* 3$^{rd}$ ed. McGraw Hill.

[13]   P. M. Hahn. (1962). Theoretical diversity improvement in multiple frequency shift keying, IRE *Trans. Communications System*, vol. CS-10, pp. 177–184.

[14]   W. C. Lindsey. (1964).  Error probabilities for Ricean fading multichannel reception of binary and N-ary signals,  *IEEE Trans. Inf. Theory*, vol. IT-10, pp. 339–350.

[15]   P. J. Crepeau. (1992).  Uncoded and coded performance of MFSK and DPSK in Nakagami fading channels,  *IEEE Trans. Commun*., vol. COM-40, March 1992, pp. 487–493.188

**Mohammad Waris Abdullah** is a lecturer and researcher at the Department of Communications & Network Engineering- King Khalid University, Kingdom of Saudi Arabia. Mr. Abdullah holds a masters degree in Wireless Communications from University of Southampton, UK. He works in the area of communications – particularly related to modulation techniques, Wireless Sensor Networks and Body Centric Communications. His research area also includes antenna design for the above mentioned areas.

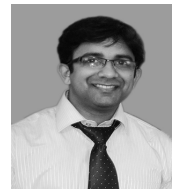

**Nazar Waheed** is a lecturer and researcher at the Department of Communications & Network Engineering- King Khalid University, Kingdom of Saudi Arabia. Mr. Waheed holds a masters degree in RF and Wireless Communications from University of Leeds, UK. His research interests are in WSN communication and WSN protocol design.

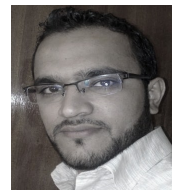